\documentclass[%
 reprint,
 amsmath,amssymb,
 aps,
]{revtex4-2}

\usepackage{graphicx}% Include figure files
\usepackage{dcolumn}% Align table columns on decimal point

\usepackage{gensymb}
\usepackage{amsmath}

\usepackage{siunitx}

\usepackage{mathtools}

\usepackage{nicematrix}

\begin{document}

\preprint{APS/123-QED}

%\title{Manuscript Title:\\with Forced Linebreak}% Force line breaks with \\
\title{When sound wave meets the neutrino anomaly}% Force line breaks with \\

\author{Jae Jun Kim}
\affiliation{%
South Carolina Department of Education, Research and Data Analysis, Columbia, SC, The United States
}%


%\author{Another Author}
%\affiliation{
%\affiliation{
 %Third institution, the second for Charlie Author
%\affiliation{%


\date{\today}% It is always \today, today,

\begin{abstract}
%An article usually includes an abstract, a concise summary of the work
%covered at length in the main body of the article.

%\begin{description}
%\end{description}
%Abstract here.
We propose that the sound wave coming to the inner side of Miniboone detector could be one of the sources for our having the neutrino anomaly in the experiment.  We start with presenting a rough estimate for the size of the energy associated with the sound wave coming into the detector, the size of the loss of energy associated with the sound wave as it travels down to the target medium due to their gravitationally interacting with the sound wave in a classical sense.  After that, we describe that the neutrino anomaly could be due to the sound wave interacting with the detector material under the pressure due to the presence of the mineral oil and their producing phonon-induced electrons via a process such as the thermionic emission, which may lead more events to be identified as electron-like events in the experiment.  We also address that the sound wave may scatter with the electrons produced from the electron-photon shower.
\end{abstract}

\maketitle

%\tableofcontents

%\section{\label{sec:level1}First-level heading:\protect\\ The line
%break was forced \lowercase{via} \textbackslash\textbackslash}

%\section{Summary}
%\label{summary}
Since we empirically confirm that neutrinos do oscillate, physicists have been working hard to uncover how our nature works in the neutrino sector.  However, as always, there are measurements that does not fit under a set of theoretical framework.  The neutrino anomaly in Miniboone experiment \cite{reference1} could be classified as one of them.  In this study, we propose that the sound waves with gravitational mass \cite{reference2} created in the electronics room and coming into the inner side of the Cerenkov detector, could be a candidate that can cause what we have observed as the anomaly, the excess, in the experiment.

In essence, what Miniboone reported as the anomaly was the estimated size of neutrino quasi-elastic signal events in the region of $< 0.5$ $GeV$ of the incoming neutrino energy does not match with the data, due to the electron- or photon-like excess \cite{reference5} in the energy region. 
After the result was reported, many different theories have been proposed \cite{reference11} to address the issue presenting some potential resolutions, including \cite{reference6}.  A combination of the resolutions have been considered when addressing the issue.  In addition, it was pointed out that more caution may need to be taken when performing their systematic study, which may reduce the confidence level \cite{reference7} for the result too.  All the studies so far have led us to reduce the difference between our projection, Monte Carlo, and the data, but their being different $\sim$ 12$\sigma$ of statistical significance \cite{reference1} still has been haunting us.

In this letter, we present an idea that could be taken a part of the resolutions from a classical point of view.  Sound wave, which was shown to carry gravitational mass in the second order \cite{reference2}, that has been created from the racks in the electronics room that is located nearby the access portal, the top of the detector and continuously coming into the inner side of the Cerenkov detector filled with mineral oil as a part of Miniboone experiment, could be one of the main causes for our seeing the anomaly, as the wave interacting with the detector material that is made of steel, producing phonon-induced electrons in the classical sense of thermionic emission under the pressure due to the presence of the mineral oil \cite{reference16}.  We also mention that the wave could scatter with the electrons in the electron-photon shower, which may demand some additional layers of simulation for the path of light within the detector.

%We focus on the former at this moment.

%In \cite{reference2}, it was shown that sound as a wave packet does carry gravitational mass.  In addition, it was shown that the wave are not only the source but also that could be affected by the gravity.  Following their calculation, the mass that the sound carries for the ordinary equation of state is negative.  In \cite{reference2}, all the details of the calculation on how they end up with the result is described.

It was stated in \cite{reference2} that the size of the mass that is carried by sound wave is proportional to that of the energy associated with the sound and depends on the medium's equation of state as a coefficient.  In the first order, the size of mass for the sound wave can be written roughly as \cite{reference2},
%{Write it down.}
\begin{equation}
E \sim m{v^2}
\end{equation}
, where $m$ is the gravitational mass, $E$ is the energy carried by sound wave and $v$ is the velocity of the sound wave in the medium.  Having such a relationship allows us to do the calculation of the scattering of the sound wave as it travels in the target medium, in a classical sense.% as not introducing effective field theory.

Coming back, if the power associated with the sound wave created in the electronics room above the detector can be measured, we can calculate the size of the mass, estimate the size of the attenuation and scattering rate as it travels to the detector, estimate the size of the loss of energy of the wave due to its interacting with the target medium.  After showing that all the contribution being negligible in terms of the fractional loss of energy, we then consider a possibility of the phonon-induced interaction as a part of the thermionic emission in the classical sense, with the mineral oil being the cause for the pressure on the surface of the detector medium.  The scenario is mathematically simple.  There are many uncertainties to be considered, but our approach can give us some idea regarding the possibility of the sound wave being a candidate for the cause of the anomaly in the experiment.

We start with the sound wave created in the electronics room \cite{reference13}, where we have all the electronic devices being placed nearby a top polar cap of the detector.

Due to the energy associated with the sound wave is not known at this moment, we do some estimation based on some of studies conducted in the acoustic research and some commonly known information.  What we have in the room is all the electronics needed for the experiment.  Assuming that the size of the electronics room is $\sim$ 200 $m^3$ and that of the aperture through which the electronics room is connected to the detector is small, $\sim$ 1 $m^3$ \cite{reference3}, compare to the size of the room, the wave created by the electronic devices resonate within the room and a part of the waves as a wave packet can enters into the detector, either by the top polar cap, or transmitted through the materials nearby the cap.  We assume that the frequency for the wave created in the room is $\sim $ 100 $Hz$, which is reasonable for the ones that is created by the fans installed to cool down the temperature of the racks. %and the intensity of the sound wave could be reduced by a factor of $\sim$ 10 due to the size of the aperture of the polar cap with respect to that of the electronics room.  In other words, the pressure due to the sound wave may need to be reduced by the factor.
Note that the detector can be considered as a large Helmholtz resonator \cite{reference15} with the size of the frequency. %but due to that the size of the aperture is larger than or similar to the wavelength associated with the sound wave in the air, we may not consider the detector as the resonator for now.

%We can start with the size of the frequency associated with a sound wave that are usually dissipated and resonate by devices in an electronics room to by in the order of 100 Hz.
%
%\begin{equation}
%f \sim 100 Hz
%\end{equation}
%.  For the cavity, the size of the frequency associated with a sound wave can be calculated.  They are,
%
%{Write it down.}
%

In essence, what makes our idea as something to be considered is that the sound wave gets keep generated and coming down to the inner side of the detector due to all the electronics in the room running continuously.  The experiment operates all the time.  That being said, if the attenuation length of the amplitude for the sound wave is long enough to get to the center of the detector, given the shape of the detector being a sphere, there is a possibility of the sound wave being one of the causes for the anomaly as it reside there in a form of standing waves.%as a wave packet with a certain resonance frequency within the detector over the time although additional simulation study may be needed later.

Conservatively, we can say that the average size of amplitude associated with the sound wave created in the electronics room to be $\sim$ 50 $dB$, which is about the size of a normal conversation.  Assuming that the electronic noise that is created mainly due to the cooling fans being under operation all the time and that the size of the intensity of the wave may get reduced by $\sim$ 100 due to the size of the top polar cap with respect to the size of the electronics room.
%\begin{equation}
%\end{equation}
With all that, we can roughly estimate the power associated with the sound wave to be  $\sim$ 1 $GeV$ per second.  It may need to get reduced down further depending on the angular distribution of the sound wave entering the mineral oil though.   
%{Write it down.}
%\begin{equation}
%E \sim 100 GeV
%\end{equation}
%

Following the equation of state of the sound wave in the air in Equation 1 \cite{reference2} and the intensity of the sound wave, the mass density, the mass associated with the sound wave coming into the detector site per second can be estimated as,
\begin{equation}
m_s \sim 500 \hspace{1mm} eV
\end{equation}
%{Write it down.}
.  Note that the mass is calculated based on the speed of sound in the mineral oil.  In essence, based on Equation 1 and 2, we can think of the sound wave with 1 $GeV$ of energy with 500 $eV$ of mass is coming into the detector in every second.  %Having such allows us to do the scattering rate in a classical sense later. 

Since sound wave can gravitationally interact with other entities \cite{reference2}, first we estimate the loss of energy due to the attenuation and then the scattering with the target medium to see how large they are.  %In any case, they having gravitational mass lead us to consider their classically hitting the detector material, which is made of aluminum.  %In short, we want to see whether it is negligible or not.

%Assuming that the sound wave is going to be uniformly distributed within the inner side of the detector after a certain time being elapsed, we consider a few different scenarios.  First, we simply estimate the size of the energy loss for the incoming neutrinos scattering with the sound wave.

%The mean free path for the sound wave is about the radius of the detector, which is $\sim$ 500 cm.
Starting with the attenuation, due to the detector is filled with mineral oil, the attenuation \cite{reference9} of the amplitude of the sound wave coming to the detector needs to be taken into account.  The degree of the attenuation depends on the frequency of the wave and the density of the medium,
\begin{equation}
\frac{A}{A_0} \sim e^{-\alpha\cdot x}
\end{equation}
, where $A_0$ is original amplitude of pressure, $\alpha$ is the attenuation coefficient and $x$ is the distance the wave travels \cite{reference9}.  %The coefficient can be written as,
%\begin{equation}
%\alpha \sim \frac{c\cdot\omega^2}{\rho_m}
%\end{equation}
%, where $c$ is dynamic viscosity coefficient, $\omega$ is angular frequency of the wave and $\rho$ is the density of the medium \cite{reference9}.
Given that $\alpha \sim$ 0.3
for the wave with 1 $MHz$ \cite{reference12} per $cm$ and the linearity of the coefficient and the average travel distance is $\sim$ 500 $cm$ the radius of the detector,
\begin{equation}
\alpha \sim 0.01
\end{equation}
, for the sound created in the electronics room.  With that, the loss of energy of the sound per second is $\sim$ 3$\%$.  When the wave travels the entire detector, it is going to lose $\sim$ 6$\%$ of its energy.

In practice, the amplitude needs to be reduced further depending on how we consider the attenuation while the wave coming into the cap or nearby the cap.  Due to $\alpha$ $\sim$ 100 in the air, it could almost completely attenuate the wave.  For instance, if we assume that the wave travels $\sim$ 100 $cm$ on average in the air before it enters the detector, then the amplitude needs to be reduced down further by $\sim$ 90$\%$, which has more impact than that travels in the target medium.  We do not know the size of the average travel distance for now and how the sound created from the racks in the electronics room is going to interfere.  Such could be tested with a device that measures decibel though.

%For now, we may estimate the size of the energy of sound wave coming into the detector per second to be $\sim$ 1 $GeV$ and their mass $\sim$ 500 $eV$.
%For instance, if we consider only Helmholtz resonance frequency of a rigid cavity \cite{reference15}, the size gets smaller significantly though.

%\begin{equation}
%m_s \sim 800 \hspace{1mm} eV
%\end{equation}
%.  With that in mind, the average energy density associated with the sound wave need to be adjusted due to the attenuation while it travels in the mineral oil.

%Note that the viscosity coefficient and the density of the medium could be a factor for our not seeing the anomaly for the Cerenkov experiment with water as medium \cite{reference10}, while we do in Miniboone.  %Then we have, 

%{Write it down.}

Do note that the sound wave carries a negative mass in the normal state \cite{reference2} so they gravitationally interact with the target medium in a repulsive manner. %for massive neutrinos or any other matters.
Given that, what we can do, given the equation of state, is to consider the wave interacting with the detector medium and then scale it down to the gravitational scale, to see whether it matters or not due to the two have a same form of the scattering amplitude.

The calculation for the energy loss due to Coulomb interaction while a particle with an electric charge travels in a medium can be found in a literature including \cite{reference8}.  The energy loss, in case of low velocity, where $c \gg v$, can be written as,
\begin{equation}
%\begin{split}
%\delta E &=  CN \cdot PE \cdot KE \cdot IP \cdot NT \cdot DS 
%\\
\delta E \sim 2\pi \cdot \frac{z^2e^4}{16\pi^2\epsilon^4} \cdot \frac{1}{mv^2} \cdot ln \hspace{1mm}\frac{2mv^2}{B} \cdot Z\cdot\frac{N}{A} \cdot \rho \cdot \delta x
%\end{split}
\end{equation}
, where $m$: mass of the electron, $B$: binding energy, $\delta x$: the travel distance and,
\begin{equation}
n = Z\cdot\frac{N}{A} \cdot \rho
\end{equation}
, is number of electrons per unit volume.
%two letter terms in the first line of the equation stand for,
%$CN$: Constant, $PE$: Electric potential, $KE$: Kinematic energy, $IP$: Impact parameter, $NT$: Number of electrons per unit volume, $DS$: Distance traveled.

%The goal here is to calculate the order of magnitude and since the sound wave with mass interacts with matters gravitationally, one thing we can do is to calculate the size of the scattering if the wave with the incoming neutrino assuming it is the case of Coulomb scattering, and then scale it down to the gravitational scale, to the first order. 
%For that, we do some substitutions.  Assuming that neutrinos move at $c$ and not considering the minimum ionization-like energy and that the loss of energy is constant as a function of the distance, we have,
When we do the same but for the gravitational scattering of the wave to the target medium by replacing some of the variables in the equation, taking that the equation of state for the gravitational scattering has a same form into account.  Then we have,
\begin{equation}
%\delta E = 2\pi \cdot G^2M^2m^2 \cdot \frac{1}{mv^2} \cdot ln \hspace{1mm}{2mv^2} \cdot Z\cdot\frac{N}{A} \cdot \rho \cdot \delta x
\delta E \sim 2\pi \cdot G^2M^2m^2 \cdot \frac{1}{mv^2} \cdot ln \hspace{1mm}\frac{2mv^2}{B} \cdot Z\cdot\frac{N}{A} \cdot \rho \cdot \delta x
%\delta E \sim G^2 \cdot M \cdot m
\end{equation}
, where $M$ is the mass of the target and $m$ is the mass of the sound wave and assumed that the sound travels with $v$ the whole time.  Assuming the impact parameter $\sim$ 1 and given that,
%Scattering of the neutrino gravitationally with the sound wave is going to cause some loss of energy for.
\begin{equation}
m \sim 500 \hspace{1mm}{eV},
\hspace{1mm} v \sim \hspace{1mm} 1500 \hspace{1mm} \frac{m}{s}, \hspace{1mm} \delta x \sim 500 \hspace{1mm} cm
\end{equation}
%, and
%\begin{equation}
%N \sim 6 \times 10^{23} \hspace{1mm}, M = \frac{4}{3}\pi r^3\rho
%\end{equation}
%, and with mass density of the mineral oil,
%\begin{equation}
%\rho_{s} = m \cdot Z\cdot\frac{N}{A} \cdot \rho
%\end{equation}
, we estimate the mean value of the loss of the energy associated with the sound wave to be in the order of,
\begin{equation}
\delta E \sim 10^{8} \hspace{1mm}eV
\end{equation}
per second, where it is assumed that it follows the Gaussian curve \cite{reference8} and the speed of sound while it travels is approximately a constant.  Note that the loss of energy is estimated as they scatter with the entire target medium in a classical sense.  Based on the result, we see that the loss due to the target mass in the detector medium is going to be negligible as it crosses the detector, which is $\sim$ 10 $m$.  The sound wave, once it enters, it going to cross the entire detector at least a few times.
%small compare to the central value of the energy.  Given the size, as expected, the scattering may not be one of the main reasons for the sound wave losing their energy as it travels in the oil.

%In other words, once the wave survives and enters the detector, it is going to survive in the order of a second, which is long enough for the wave to cross the detector $\sim$ 100 times.

%through the air as the medium and enter into the mineral oil, 

%particularly with the reason that the wave forms a standing wave within the detector as a Helmholtz resonator.
%Considering its scattering with the medium of the target, the loss is not worth to be considered.

%However, that being said, due to the wave has a such a small scattering amplitude, doubly suppressed by $\sim G^2$ in the expression, and its not interacting via other fundamental forces but only by gravitational forces, the next scenario that we can consider is the sound wave being a part of the target medium itself, in a classical sense.
%while it does not get fully attenuated.  %In other words, we can think of something that indirectly induces other interactions.  For instance, electron-phonon scattering \cite{reference14} could be one of them.

%Due to the sound wave has a negative mass in normal equation of state, 
%The contribution of the wave to the target mass in the detector medium is likely to be negligible too, given the target density of the mineral oil.
However, one scenario that we may need to consider is the wave traveling to the boundary of the detector, which is made of steel, interacting with the detector material, not the target medium, via a phonon-induced interaction.  Based on our calculation, the attenuation length is long enough on average for the wave to get to the opposite side of the detector, especially due to the shape of the detector being a Helmholtz resonator.  If the wave can induce an interaction of the phonon with the valence electrons on the detector material, classically, producing knock-out electrons coming into the inner side of the detector.  Such could potentially lead more muon-like events to be identified as electron-like as opposed to their being identified as a background event.  Or, the electron-like events with very low energy can be identified as that with a higher energy.

When we say the phonon interacting with the electrons in steel, it is not about the electron-phonon interaction in solid state, where the size of the energy exchange is $\sim$ $E_F$, but more like as a classical version of a phonon-induced effect, one that is similar to a thermionic emission in the acoustic welding \cite{reference17}.  In other words, the phonon with gravitational mass is going to scatter with the surface of the detector material, and as the energy gets accumulated to the binding energy of the valence electron, it shoots out the electrons that corresponds to the energy of the phonon.  It is different from the photo-electric effect in that the classical accumulation of energy hitting on the surface and causing some valence electrons on the very surface to pop out.  The photo-electric-like energy that corresponds to the sound wave is anyway going to be too small since it is
$\sim$ $10^{-10}$ $eV$
, when we consider either the resonance frequency or the frequency of the sound wave generated in the electronics room.  However, again, the sound wave can interact with the surface medium in the classical sense.  Furthermore, the presence of the mineral oil is going to provide the pressure to the detector material, especially on the bottom portion of it.  Taking the density of the oil and assuming that the average height of the oil filled in the detector to be $\sim$ 500 $cm$, the pressure could be $\sim$ 50,000 $Pa$
, which could be in the order for that in the industrial acoustic welding to initiate the process.  Once the electrons are produced, they can drift into the detector.

In the scenario, we may need to consider that the surface of the main detector is painted thus the bare steel, which is what the detector is made of, is not what the sound wave is going to meet but the painted material.  Such a layer could raise the temperature of the surface of the steel even faster and absorb more sound wave thus it makes hard to predict how much of the sound amplitude is going to be reflected and how much being absorbed.  %Such could be answered with some detailed simulations.
The work function for the valence electrons for the detector material is $\sim$ 4 $eV$ \cite{reference14}.  %At this moment, it is hard to say at which rate the knock-out electrons are going to be produced and how the energy spectrum is going to look like.

In addition, as a last scenario, the sound wave can scatter with the electrons produced in the electron-photon cascade, which can happen for the signal event.  We do not know how the path for the sound wave look like at this moment, particularly due to the presence of the photo-multiplier tubes nearby the access portal and due to that of the supporter.  Depending on how the path like, it could cause an addition layer of multiple scattering to the path of the electrons in the cascade.

You may wonder why the scenario is applicable for an experiment such as Miniboone but not for the others.  The answer might rest on the fact that mineral oil is where the drifted electrons produced in the thermionic emission can take place, and the shape of the detector being a sphere, a suitable shape for sound wave to resonance as forming a standing wave, thus last longer as getting less attenuated with smaller frequency, in combination of a the cylindrical shape of the polar cap having a short length thus the attenuation due to its traveling in the air can be mitigated.  Water as a medium with polarity, being used as a medium in other experiments such as Super-Kamiokande might be a reason for the anomaly not being present in there.  %Electrons are emitted, if any, as the sound initiating the thermionic emission and they are going to wander within the detector.% due to the oil having no conductivity at all, whereas water does have many conduction electrons.
%especially nearby the boundary region due to the pressure.  %If we take a wave packet approach, then it could be argued that the resonance of the incoming sound wave as a wave packet interfere a lot more in water than that in the mineral oil.
%In addition, sound having negative mass lead them not to interact with other matters further due to their being repulsive against all other matters with positive masses, thus having a longer time to survive within the detector medium.  %Or the detector being a shape of sphere for the case of Miniboone could be a reason.

Note: You may also wonder whether we can isolate the effect due to the sound wave while running all the electronic devices when the beam is not coming.  The effect certainly could be identified by studying the abnormal behaviors in the response of the detector including the gain of the photomultiplier tubes.  However, if the effect is below the threshold of the gain, below the detector resolution, above which the response can be identified, it may not show up when running the detector without the beam.  In such a case, with the electrons due to the emission being almost uniformly scattered within the detector, could cause some muon-like events below the energy threshold of the measurement, to be identified as the signal events.

One of the tests we can do to see if that really is the case is to create a prototype settings, where we have the detector medium in a much smaller scale but add a source where the sound wave can be replicated and then see how much change we can observe in terms of the spectrum of the incoming neutrino to the detector.  In addition, Miniboone collaboration conducted a study to simulate the energy response of ionized electrons produced within the detector in the mineral oil \cite{reference3}.  The work done there can be used again as a basis when conducting the test.  Or, we can start simply by measuring the decibel of sound within the detector, particularly nearby the top polar cap and in the middle of the detector, while it runs.

In summary, it is a bit radical idea.  The key point is this: 
How large is the size of the energy associated with the sound wave entering the mineral oil going to be, after being heavily attenuated in the air while it travels and after either going through or reflected in the top portion of the detector where the photomultiplier tubes are present?  If, with some chance, some survives, are they really going to initiate the thermionic emission?  If the answer to the two questions are yes, then it may be worth to consider the scenario as a resolution, partially or fully, due to the anomaly has been haunting us for a while.
%\begin{acknowledgments}
%We wish to acknowledge the support of the author community in using
%REV\TeX{}, offering suggestions and encouragement, testing new versions,
%\end{acknowledgments}
\begin{acknowledgements}
The author truly thanks his family for all their support.
\end{acknowledgements}

%\nocite{*}

%\begin{thebibliography}{}
%\bibitem{reference1}



\end{document}